\documentclass[%
 reprint,
 amsmath,amssymb,
 aps,
]{revtex4-2}

\usepackage{graphicx}
\usepackage{dcolumn}
\usepackage{bm}
\usepackage{braket}
\usepackage{color}
\usepackage[version=3]{mhchem}
\usepackage{mathtools}
\usepackage{newtxtext}
\usepackage[varg]{newtxmath}
\usepackage[normalem]{ulem}
\usepackage{lipsum}
\usepackage{comment}
\usepackage[final]{hyperref}
\hypersetup{colorlinks=true,linkcolor=blue,citecolor=blue,urlcolor=blue}

\begin{document}

\preprint{APS/123-QED}

\title{Quasi-particle dynamics in quasi-periodic Ising model \\ with temporally fluctuating transverse fields}

\author{Kohei Ohgane$^1$, Yusuke Masaki$^1$, and Hiroaki Matsueda$^{1,2}$}
\affiliation{
 $^1$Department of Applied Physics, Graduate School of Engineering, Tohoku University, Sendai 980-8579, Japan\\
$^2$ Center for Science and Innovation in Spintronics, Tohoku University, Sendai 980-8577, Japan 
}%
\date{\today}

\begin{abstract}
We study quasi-particle dynamics in a quasi-periodic Ising model with temporally fluctuating transverse fields.
Specifically, we calculate the dynamical exponents of the standard deviation of a quasi-particle spreading under a field chosen randomly from binary values $\pm h$ at every time interval.
We find that the short-time behavior of the dynamical exponents depends on the interval of the temporally fluctuating fields.
We also reveal how the quasi-particle dynamics affects the relaxation of spin-spin correlation functions. 
The dynamics can be explained via the overlap between the eigenvectors of a Hamiltonian with $\pm h$.   
\end{abstract}

\maketitle
\section{Introduction}
Anderson's pioneering research has opened a door to analysis of localization phenomena in randomly disordered systems \cite{Anderson}.
Through various studies on Anderson localization, 
much knowledge has been obtained for cases of one-particle problems \cite{Abrahams,Igloi_randomTFIM,Evers}.
Recently, localization with many-body interactions has attracted much attention in the context of thermalization problems in isolated quantum systems \cite{Nandkishore,Alet,Abanin}.
Such localization is referred to as many-body localization (MBL). 
MBL is also an important phenomenon in quantum information engineering because an MBL phase can retain local information at an initial state during time evolution.
In this way, MBL provides a vast research stage across various fields of physics.

Anderson localization has also been studied in the field of quantum walks,
which were originally introduced as quantum versions of classical random walks \cite{Aharonov}.
The propagation of walkers in quantum walks is different from that in classical random walks.  
While a random walk exhibits a diffusive dynamics that is characterized by a standard deviation of the probability distribution, given as $\sigma \sim \sqrt{t}$, a walker in a quantum walk propagates ballistically as $\sigma \sim t$. 
Such ballistic dynamics is obtained in homogeneous cases.
The dynamics in a quantum walk is strongly affected by spatial and temporal disorder.
Spatial disorder localizes quantum walkers as $\sigma\sim 1$ \cite{Joye, Konno_1, Konno_2, Obuse, Ahlbrecht_spatio_1}.
On the other hand, temporal disorder leads to diffusive dynamics as $\sigma \sim \sqrt{t}$, regardless of whether spatial disorder exists \cite{Abal, Chandrashekar, Kosik, Shapira, Romanelli, Leung, Ahlbrecht_temopral, Montero,Burrell}.
Because quantum walks are almost equivalent to the dynamics of excitations in tight-binding models,
results for quantum walks are useful in understanding the localization phenomena and diffusive dynamics in disordered systems.

Quantum walks are formulated in two different ways: as a discrete-time quantum walk (DTQW), or a continuous-time quantum walk (CTQW) \cite{Mulken}.
The difference between these formulations is in how the system evolves in time.
The time evolution in a DTQW is implemented by the product of two unitary operators, a coin operator and a displacement operator.
On the other hand, the time evolution operators in CTQWs are given by solutions of the Schr\"{o}dinger equation.
In this way, a CTQW is closer to the formulations of condensed matter physics and the dynamics is related to non-equilibrium relaxation as compared with a DTQW. 
Sachdev first showed that the relaxation of observables can be obtained from the classical trajectories of quasi-particles \cite{Sachdev}.
These trajectories are estimated from the time evolution of the standard deviation of the probability distributions of the quantum walkers in CTQWs.
This framework has been applied to integrable and disordered systems, and it can successfully estimate local quantities and correlations \cite{Rieger,Igloi_quasicrystal}. 
Later, Ro\'{o}sz \textit{et al.} analyzed the quasi-particle dynamics in a one-dimensional transverse field Ising model (TFIM) whose transverse field fluctuates in the time domain \cite{Roosz}.
The main characteristic parameter of the randomly fluctuating field is its time interval $\tau$. 
They showed that the standard deviations exhibit diffusive or super-diffusive behaviors given by $\sigma \sim t^{1/z} ~(0.5 \leq 1/z < 1)$, depending on the interval $\tau$.
However, they did not consider spatial disorder, and in this paper, we reveal whether a super-diffusive dynamics can be obtained even under spatial disorder.

\begin{figure}[t]
\centering
\includegraphics[width=7cm]{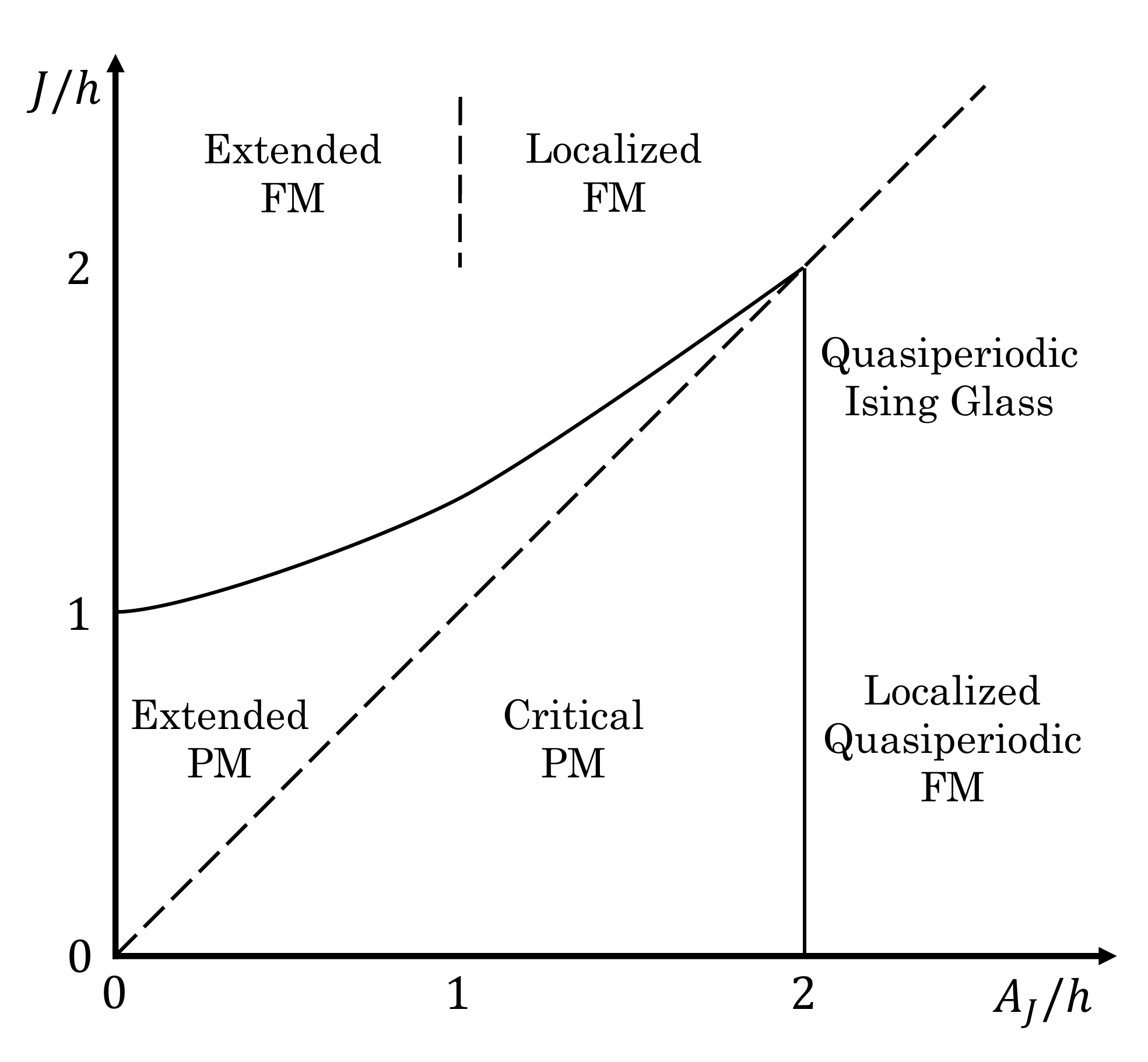}
\caption{Phase diagram of the QP-TFIM. 
The complexity is due to the various symmetries in individual parameter regimes.
The abbreviation PM and FM indicate paramagnetic and ferromagnetic, respectively.
The diagram was obtained by A. Chandran \textit{et al.}}
\label{fig:phase_diagram_QPTFIM}
\end{figure}

Like random disorder (RD), quasi-periodicity (QP) also leads to localized eigenstates.
The potential or hopping parameters of QP systems are spatially modulated with incommensurate periods.
Although more than 40 years have passed since Abzel, Aubry and Andr\'{e} reported localization in QP systems \cite{Abzel,Aubry_Andre}, 
fewer studies have been conducted on these systems than on RD systems.
Unlike RD systems, QP systems have delocalized eigenstates even in one dimension. 
In addition, QP systems at critical points have fractalities in their eigenspectra and eigenfunctions \cite{Ostlund,Hofstadter,Ketzmerick},and they exhibit anomalous diffusion \cite{Hiramoto,Piechon,Setiawan,Roosz_quasiperiod}. 
While QP systems provide these rich properties, there is not much choice in QP models:
the Aubry-Andr\'{e} model is the main stages of studies on localization in QP systems \cite{Ganeshan,Biddle_pow_hop,Biddle,Iyer}.
Recently, another QP model, the quasi-periodic transverse field Ising model (QP-TFIM), has come to be studied \cite{Chandran,Crowley,Divakaran}. 
The QP-TFIM has the complex phase diagram depicted in Fig. \ref{fig:phase_diagram_QPTFIM}. 
From the viewpoint of localization, the phases are roughly categorized into three types: extended, localized, and critical phases.
In the extended phase, quasi-particles propagate ballistically, whereas no transport is observed in the localized phase.
The critical phase is located between the other two phases and exhibits extremely slow dynamics.
The dynamics under the static Hamiltonian of the QP-TFIM was clarified previously \cite{Divakaran}.
However, because systems are exposed to fluctuating external fields in realistic situations, it is worthwhile to reveal the dynamical properties in a QP system under fluctuation in the time domain.

In this study, we consider the quasi-particle dynamics in the QP-TFIM under temporally fluctuating fields. 
The transverse fields are randomly chosen from binary values $\pm h$ at every duration with a certain interval $\tau$. 
The way the quasi-particles spread is a key to understanding the relaxation processes, because the local observables and correlation functions can be estimated from this dynamics in a semi-classical framework.
Thus, we perform stroboscopic time evolution and average the probability distribution over realized sequences of temporally random fields.
Our calculations focus on the extended paramagnetic (extended PM) region shown in  Fig.~\ref{fig:phase_diagram_QPTFIM}, which is in contact with the critical point of the TFIM.
Our results on the quasi-particle dynamics can be summarized as follows.
Short-time behaviors depend on $\tau$, and the quasi-particle dynamics is super-diffusive for certain values of $\tau$.
However, the quasi-particle dynamics approaches to be diffusive for any interval $\tau$ in the long time scale.
We also calculate the relaxation dynamics of two-point spin-spin correlation functions. 
The exponents in the relaxation of these correlation functions is consistent with the dynamical exponents of the quasi-particles.
Such dependence of the dynamical exponents on $\tau$ can be explained by the overlap between the eigenvectors of the Hamiltonian with $\pm h$.
For the TFIM, the overlap has a simple structure, which leads to nearly ballistic dynamical exponents for certain $\tau$.
On the other hand, finer structures appear as the strength of the quasi-periodic spin-spin couplings increases, which leads to a diffusive dynamics for any $\tau$.

The rest of our paper is organized as follows.
In Sec.~II, we introduce the details of the model and its formulation.
In Sec.~III, we show the numerical calculations for the quasi-particle dynamics and spin-spin correlation function.   
In Sec.~IV, we review the theory introduced in Ro\'{o}sz's paper and extend it to the QP-TFIM.
Finally, we summarize our study in Sec.~V. 

\section{QUASI-PERIODIC ISING MODEL WITH TEMPORALLY FLUCTUATING FIELDS}
\subsection{Hamiltonian}
The Hamiltonian of the quasi-periodic Ising model with temporally random transverse fields is written as
\begin{gather}
\hat{H}(t)=-\frac{1}{2}\sum_{j=1}^{L-1}J_{j}\hat{\sigma}^{x}_{j}\hat{\sigma}^{x}_{j+1} -\frac{1}{2}h(t)\sum_{j=1}^{L}\hat{\sigma}^{z}_{j},\label{eq:Hamiltonian_QPTFIM} \\ 
J_{j}=J + A_{J}\mathrm{cos}\Bigl(Q( j + 1/2 ) \Bigr),\nonumber
\end{gather}
where $\hat{\sigma}_{j}^{\alpha} ~ (\alpha = x,y,z )$ are Pauli matrices at site $j$, 
$J_{j}$ represents the quasi-periodic spin-spin coupling between sites $j$ and $j+1$, 
and $Q$ denotes an irrational number given as the golden ratio: $2\pi\times(\sqrt{5}+1)/2$.
This system is in the open boundary condition.
The system size $L$ is set large enough that the edges of the system do not affect the dynamics.   

In this paper, we consider transverse fields $h(t)$ that are spatially uniform but temporally fluctuating.
The field is randomly chosen from binary values $\pm h$ at every discrete time $t_{n}=n\tau$, where $\tau$ is the discrete time interval.
Within the duration defined by $(t_{n-1},t_{n}]$, $h(t)$ takes a constant value.

This QP-TFIM has the complex phase diagram depicted in Fig.~\ref{fig:phase_diagram_QPTFIM}.
The phases are determined by the uniform component $J$, the quasi-periodic component $A_{J}$ of the spin-spin coupling, and the static transverse field $h$. 
In this study, we aim to extend Ro\'{o}sz's work for the TFIM to the QP-TFIM.
Hence, we focus on the extended PM region ($J=1, h=1, 0 < A_{J} < 1$), which connects with the critical point of the TFIM as $A_{J}\to 0$. 
In this parameter region, the eigenfunctions are spatially extended and quasi-particles propagate ballistically with time.

\subsection{Formulation}
In this section, we formulate the stroboscopic time evolution with respect to the time interval $\tau$.
The time-dependent Hamiltonian has fields whose sign randomly changes at every interval $\tau$.
However, the Hamiltonian can be treated as a static Hamiltonian in the time domain between the intervals.
In this case, the static Hamiltonian is used to construct the time evolution of the time-dependent Hamiltonian.

As with the TFIM, the Hamiltonian of the QP-TFIM can be written in a quadratic form of Majorana fermions.
The operators of the Majorana fermions are defined with Pauli operators as 
\begin{eqnarray}
\begin{cases}
\hat{\gamma}_{2i-1} = \Biggl( \prod_{j<i} (-\hat{\sigma}^{z}_{j}) \Biggr)\hat{\sigma}^{x}_{i} \\
\quad \hat{\gamma}_{2i} = \Biggl( \prod_{j<i} (-\hat{\sigma}^{z}_{j}) \Biggr)\hat{\sigma}^{y}_{i} 
\end{cases},
\end{eqnarray}
which satisfies the anti-commutation relation
\begin{eqnarray}
\{ \hat{\gamma}_{i}, \hat{\gamma}_{j}\} =2 \delta_{i,j} \quad ( i=1,\cdots,2L).
\end{eqnarray}
The static Hamiltonian $\hat{H}_{\pm}$ can be written as
\begin{eqnarray}
\hat{H}_{\pm} = \frac{1}{4}\sum_{i,j=1}^{2L} \hat{\gamma}_{i}[\mathsf{H}_{\pm}]_{ij}\hat{\gamma}_{j},  \label{eq:Hamiltonian_Majorana}
\end{eqnarray}
where the indices $\pm$ represent the signs of the transverse fields $\pm h$.

Here, we describe the essential properties of the static Hamiltonian for explaining our results.
First, the chiral transformation changes the sign of the static Hamiltonian as
\begin{eqnarray}
\mathsf{C}\mathsf{H}_{\pm}\mathsf{C} = -\mathsf{H}_{\pm},
\end{eqnarray}
where the chiral matrix $\mathsf{C} = \mathrm{diag}[-1,1,-1,1,\cdots]$.
This leads to a property that, if $+\epsilon_{\mu}$ is an eigenenergy of $\mathsf{H}_{\pm}$, then $-\epsilon_{\mu}$ is also an eigenenergy.
Second, the Hamiltonian with the transverse field $+h$ can be converted to that with $-h$ by the operation of $\prod_{i} \hat{\sigma}_{i}^{x}$ as
\begin{eqnarray}
\Bigl[\prod_{i} \hat{\sigma}_{i}^{x}\Bigr]\hat{H}_{+}\Bigl[\prod_{i} \hat{\sigma}_{i}^{x}\Bigr] = \hat{H}_{-},
\end{eqnarray}
where $\hat{H}_{\pm}$ denotes the respective Hamiltonians with the transverse fields $\pm h$.
This relation leads to the same eigenspectra between $\hat{H}_{-}$ and $\hat{H}_{+}$, whereas the eigenvectors of $\hat{H}_{\pm}$ are different from each other and can be obtained by the operation of $\prod_{i} \hat{\sigma}_{i}^{x}$ on the eigenvectors of the other Hamiltonian.

Next, we move on to formulate the stroboscopic time evolution with respect to the time interval $\tau$.
As mentioned above, the transverse fields are chosen from the constant values $\pm h$ at every discrete time $t_n$.
The Hamiltonian is static within an individual duration $(t_{n-1},t_{n}]$.  
Thus, although the Hamiltonian is time dependent, 
its time-evolution operators from $t=0$ to $t_{n}$ can be written in products of time-evolution operators advancing time by $\tau$ as 
\begin{eqnarray}
\hat{U}(n\tau) = \hat{\mathcal{U}}_{s_{n}}(\tau)\cdots\hat{\mathcal{U}}_{s_{1}}(\tau), \label{eq:random_operation} 
\end{eqnarray}
where $s_{n}$ is the sign of $h(t)$ for the duration $(t_{n-1},t_{n}]$ and $\hat{\mathcal{U}}_{s_{n}}=\mathrm{e}^{-\mathrm{i}\hat{H}_{s_{n}}\tau}$ denotes the time-evolution operator. 

The time evolution from $\hat{\gamma}_{i}(t_{n-1})$ to $\hat{\gamma}_{i}(t_{n})$ is 
\begin{eqnarray}
\hat{\gamma}_{i}(t_{n}) = \sum_{j=1}^{2L}[\mathsf{O}_{\pm}(\tau)]_{ij}\hat{\gamma}_{j}(t_{n-1})
\end{eqnarray} 
where $\mathsf{O}_{\pm}(\tau)$ is an orthogonal matrix corresponding to the time-evolution operator with respect to the Majorana fermions.
This time evolution is obtained from solutions of the Heisenberg equations defined as
\begin{eqnarray}
\mathrm{i} \frac{d}{dt}\hat{\gamma}_{i}(t) = \sum_{j=1}^{2L}[\mathsf{H}_{\pm}]_{ij}\hat{\gamma}_{j}(t). \label{eq:eq_of_motion}
\end{eqnarray}

We next perform quantum quenches as follows.
The time evolution starts from one of the degenerate ground states in the classical Ising limit, $J\gg A_{J}, h$. 
We describe the state as $\ket{x}$, which is a product state where all spins point in the $+x$ direction.
After the above preparation, the transverse fields are suddenly switched on, and the Hamiltonian parameter after these quenches is in the extended PM region.  
In the time evolution under the Hamiltonian after the quenches, the transverse fields rotate the spins in the $x-y$ plane, which creates numerous kinks.
At the same time the kinks are created, they begin to propagate through the system. 
The Majorana fermions correspond to the creation and annihilation operators of the kinks, which can be checked by the operation of $\hat{\gamma}_{i}$ on $\ket{x}$.
Thus, after the quenches, numerous Majorana fermions are emitted from each site and propagate through the system.

To understand the non-equilibrium dynamics following the quenches, it is important to reveal how the Majorana fermion propagates through the system. 
In reference \cite{Rieger}, the dynamics of the Majorana fermion is treated in the semi-classical framework for understanding the relaxation.
In this framework, the spin-spin correlation functions $\bra{x}\hat{\sigma}^{x}_{i}(t)\hat{\sigma}^{x}_{j}(t)\ket{x}$ can be estimated from the classical trajectories of the quasi-particles.
Since the Majorana fermions roughly correspond to kinks, when an odd number of kinks cross the line from $(i,t)$ to $(j,t)$, the correlation between $\hat{\sigma}^{x}_{i}(t)$ and $\hat{\sigma}^{x}_{j}(t)$ becomes negative; that is, $\hat{\sigma}^{x}_{i}(t)$ points to the opposite direction to $\hat{\sigma}^{x}_{j}(t)$. 
By averaging the individual signs of the correlation functions over the number of kinks crossing the line, $\bra{x}\hat{\sigma}^{x}_{i}(t)\hat{\sigma}^{x}_{j}(t)\ket{x}$ can be estimated.
The negative correlations contribute to relaxations of the correlation function.
Thus, the speed of the relaxation is directly determined by how quasi-particles propagate.

The propagation of the Majorana fermions can be described by a two-point correlation function defined as 
\begin{eqnarray}
G_{i}(t) = \frac{1}{\sqrt{2}}\bra{x} \hat{\gamma}_{i}(t) \hat{\gamma}_{L}\ket{x}.
 \label{eq:def_G}
\end{eqnarray}
The physical meaning of this correlation function is the probability amplitude of the Majorana fermion propagating from a space-time point $(L, 0)$ to $(i, t)$, where the space coordinate is written in the Majorana representation.
The initial site $L$ represents the central position of the system.
At the initial time $t=0$, $G_{L}(0) =1/\sqrt{2}, G_{L+1}(0) = \mathrm{i}/\sqrt{2}$.
By using the orthogonality of time-evolution matrices $\mathrm{O}_{\pm}$, one can find that
\begin{eqnarray}
\sum_{i=1}^{2L} |G_{i}(t)|^{2} = 1 \label{eq:sum_rule}.
\end{eqnarray}
Because the Hamiltonian (\ref{eq:Hamiltonian_Majorana}) has spatial variation, the time evolution of the two-point correlation function (\ref{eq:def_G}) may depend on where the Majorana fermions are initially created.  
Fortunately, in the extended PM regime, such dependence is negligibly small after the averaging over the various realizations of the fluctuating fields.   
Hence, we focus on the case in which the Majorana fermion is created at the center of the system. 
By analogy with CTQWs, we define the probability distributions as
\begin{eqnarray}
p_{i}(t) := |G_{2i-1}(t)|^{2}+|G_{2i}(t)|^{2} ~ (i = 1,2,\cdots, L).
\end{eqnarray}

By combining Eqs.~(\ref{eq:eq_of_motion}) and (\ref{eq:def_G}), one can find the following equation:
\begin{eqnarray}
\mathrm{i} \frac{d}{dt}G_{i}(t) = \sum_{j=1}^{2L}[\mathsf{H}_{\pm}]_{ij}G_{j}(t) \label{eq:eq_of_motion_G}.
\end{eqnarray}
This equation takes the same form as that of the Majorana fermion operators.
The time-evolution operators $\mathsf{O}_{\pm}(\tau)$ advance $\{G_{i}(t_{n-1})\}$ to $\{G_{i}(t_{n})\}$ as 
\begin{eqnarray}
G_{i}(t_{n}) = \sum_{j=1}^{2L}[\mathsf{O}_{\pm}(\tau)]_{ij}G_{j}(t_{n-1}).
\end{eqnarray} 
As mentioned in Sec.~IV, the eigenvectors of $\mathsf{H}_{\pm}$ are useful for explaining the relaxation of the quasi-particle dynamics under temporal noise.

The temporally fluctuating transverse fields lead to the dependence on the random sequence. 
To obtain a result that does not depend on each realization, we take an average of $p_{i}(t)$ over $N_{\mathrm{samp}}$ realizations of temporally random transverse fields.
We denote the distribution obtained for $k$-th $(k=1,2,\cdots, N_{\mathrm{samp}})$ sequence of the field as $p_{i;k}(t)$, and we define the averaged probability distributions as 
\begin{eqnarray}
p_{i}(t) = \frac{1}{N_{\mathrm{samp}}} \sum_{k=1}^{N_{\mathrm{samp}}} p_{i;k}(t).
\end{eqnarray}

\section{RELAXATION ON TEMPORAL NOISE}
\subsection{Quasi-Particle Dynamics}

\begin{figure}[h]
\centering
\includegraphics[keepaspectratio, scale=0.37]{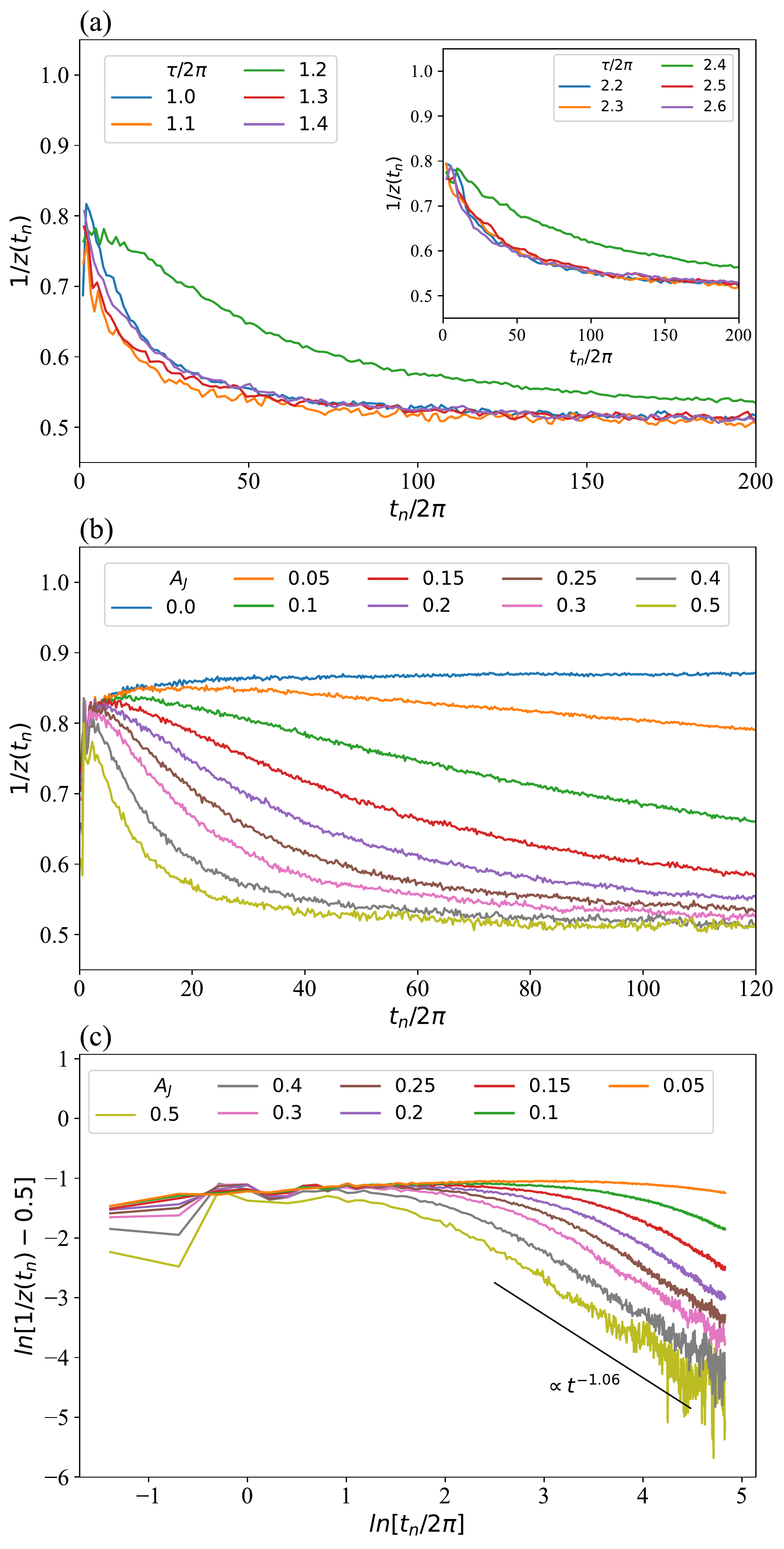}
\caption{Dynamics of inverse dynamical exponents, where all calculations were performed with $L=2048, N_{\mathrm{samp}}=10^{4}$.
(a) Dependence of the fluctuation times $\tau$ on the dynamics of $1/z(t_{n})$ at $A_{J} =0.5$. 
The inverse dynamical exponent $1/z(t_{n})$ approaches to the diffusive value $0.5$ for most fluctuation times $\tau$, except for $\tau/2\pi = 1.2, 2.4$. 
(b) Dependence of the quasi-periodic spin couplings $A_{J}$ at $\tau/2\pi = 0.25$. The dynamical exponent at $A_{J}=0$ (TFIM) does not drop to the diffusive value $0.5$. In contrast, the other dynamical exponents reach the diffusive value in the long time scale, with the speed of the decay tending to be faster with increasing $A_{J}$. 
(c) Log-log plot of the relaxation curves in (b).  
The dynamical exponents exhibit power-law decay, and the slopes of the curves become the same in the long time scale. 
}
\label{fig:zeff}
\end{figure}

To analyze how the quasi-particles propagate, we introduce the following standard deviation in the probability distribution:
\begin{eqnarray}
\sigma(t_{n}) = \sqrt{\sum_{l}p_{l}(t_{n})\Bigl(l-l_{0}\Bigr)^{2}},
\end{eqnarray}
where $l_{0} = (L+1)/2$ is the center position of the system.
As is well known for random walks (quantum walks), the time dependence of the standard deviation is obtained as $t^{1/2}~(t^{1})$.
We expect the standard deviation to take the form of
\begin{eqnarray}
\sigma(t) \sim t^{1/z}
\label{eq:sigma_t}
\end{eqnarray}
where the dynamical exponent $z$ depends on the class of the dynamics.  
Typically, $1/z = 0$ for localized systems, $1/z = 1/2$ for diffusive systems, and $1/z = 1$ for ballistic systems.
As mentioned above, we deal with the discrete time evolution here, and we calculate the dynamical exponent at the discrete time $t_{n}$ from
\begin{eqnarray}
\frac{1}{z(t_{n})} = \frac{\mathrm{ln}\Bigl[ \sigma(t_{n})/\sigma(t_{n-1}) \Bigr]}{\mathrm{ln}\Bigl[ t_{n}/t_{n-1} \Bigr]}.
\end{eqnarray}

The previous study on the TFIM investigated the dynamics of the quasi-particles under the temporally random fields with the fluctuation time $\tau$ \cite{Roosz}.
The study showed that, depending on $\tau$, the dynamical exponents took three different values corresponding to diffusive, super-diffusive and nearly ballistic dynamics. 
Note that the previous study focused on the spatially homogeneous system.

In this study, we consider the systems with the weakly quasi-periodic modulation instead of the TFIM.
The dynamics of the quasi-particles are determined by the four parameters $J, A_{J}, h$ and $\tau$, where we set $h=1$ to normalize the Hamiltonian without loss of generality.
Our calculations are performed in the extended PM region in contact with the TFIM. 
Here, we limit the discussion to the line of $J=1$ in the phase diagram (Fig.~\ref{fig:phase_diagram_QPTFIM}).
The remaining independent parameters are given by $A_{J}$ and $\tau$.

The system size $L$ is set to $2048$ to avoid effects of the system edges in the time range of the calculation. 
The quasi-particle on the center of the system at the initial time propagates outward to the edges, and then reflects at the edges. 
The reflection attributes to the boundary effect, and is not caused by the property of the bulk of the QP-TFIM.
Since our purpose is to extract the dynamical exponent in the bulk of the QP-TFIM, the reflection should be omitted from the calculation.
In addition, as the system size increases, the time till the reflection increases.   
Thus, the system size need to be prepared large enough and the time range of the calculation need to be short enough so that the size effect dose not appear. 

\begin{figure}[t]
\centering
\includegraphics[keepaspectratio, scale=0.45]{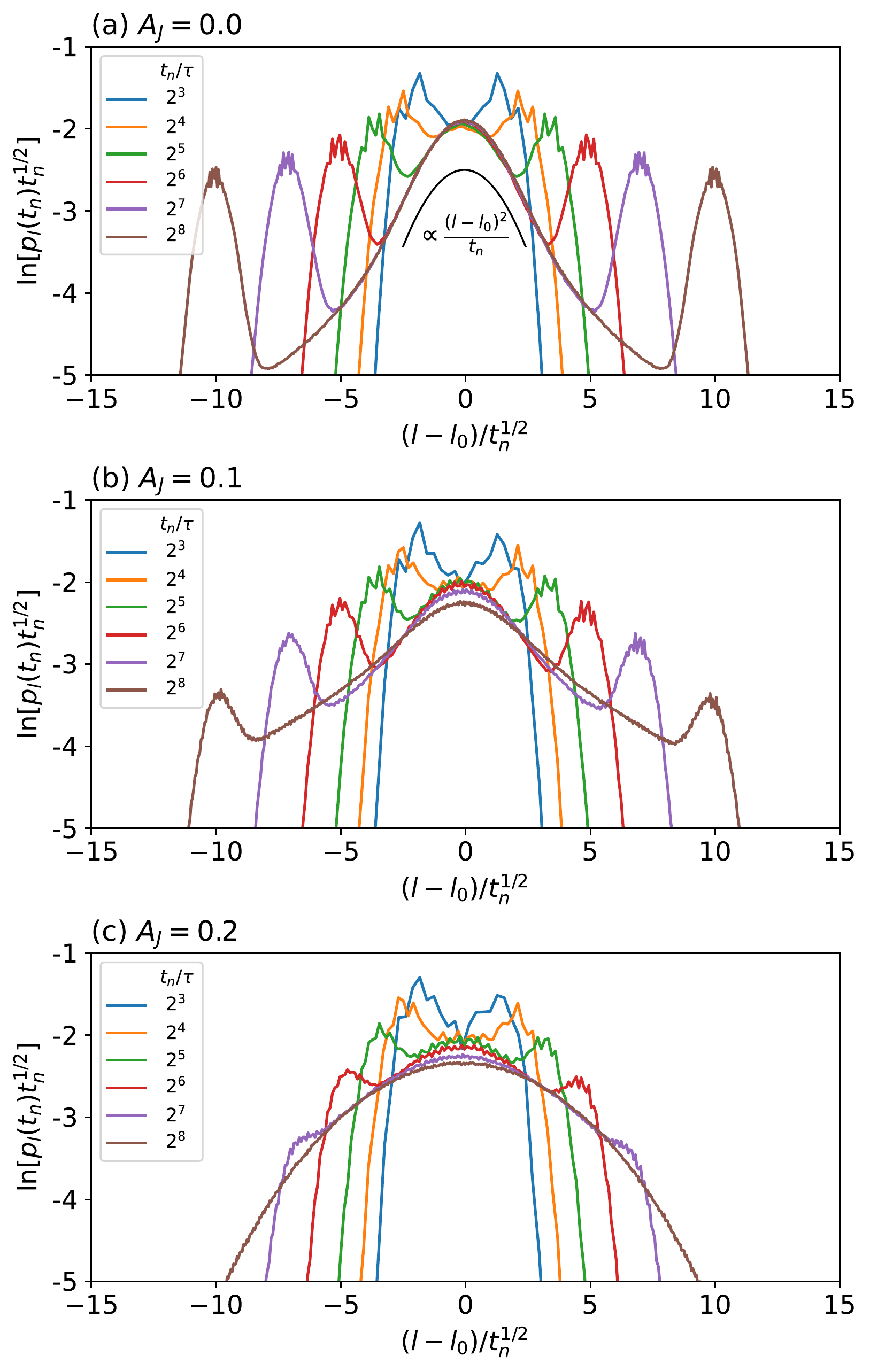}
\caption{Scaling of the shapes of the probability distributions for $A_{J} = 0, 0.1, 0.2$ (a-c, respectively). 
For $A_{J}=0$ (TFIM), two outside peaks can be seen, and the distributions are not on the quadratic curve of the diffusive scaling. 
The peaks are suppressed as $A_{J}$ increasing.
For $A_{J}=0.2, t_{n}/\tau = 2^{8}$, the distribution can be fit by the quadratic curve.}
\label{fig:scaling}
\end{figure}

We first calculate the dynamical exponent for $A_{J}/h = 0.5$ and $\tau/2\pi =1.0-1.4 $ or $2.2-2.6$ to check the dependence on the fluctuation time $\tau$, as depicted in Fig.~\ref{fig:zeff} (a). 
The dynamics following the quenches is separated into two time sectors, for short-time behaviors and long-time behaviors. 
For long-time behavior, the dynamical exponents tend to approach the diffusive value $1/z = 0.5$ for any $\tau$. 
In contrast, the dependence on $\tau$ appears in the short time scale. 
For $\tau/2\pi = 1.2$, the relaxation of the dynamical exponent is slower as compared with other fluctuation times.
The same behavior can be seen for $\tau/2\pi = 2.4$, as shown in the inset of Fig.~\ref{fig:zeff}(a).   
We choose $A_{J} = 0.5$ as a representative value, because we have confirmed that the same slow relaxations happen for other $A_{J}$ values in the extended PM region.

Next, in Fig.~\ref{fig:zeff}(b), we show how the nearly ballistic dynamics in the TFIM becomes diffusive with increasing $A_{J}$.
For $A_{J}/h =0$ and $\tau =\pi/2$, the dynamical exponent takes the nearly ballistic value $1/z \sim 0.9$ and never decays, even in the long time scale.
However, for $A_{J} \neq 0$, the dynamical exponent decays to the diffusive value.
Figure~\ref{fig:zeff}(b) shows that the decay of the dynamical exponents becomes faster as $A_{J}$ increases.
As seen in Fig.~\ref{fig:zeff}(c), the relaxation of $1/z$ in the long time scale follows a power of the time $t$. 
The slopes of the relaxation curves for all $A_{J}$ except zero in this log-log plot correspond to the same value of $\sim -1.06$ in the long time scale.
The values of the slopes are evaluated by the least square method.


It is shown that the dynamical exponent becomes diffusive in the long time scale in Fig.~\ref{fig:zeff}, but it is still unclear whether the shape of the probability distribution also becomes diffusive.
To check that, we also analyze the shapes of the probability distributions in the same setting as that used for Fig.~\ref{fig:zeff}(b).
The probability distribution for a diffusive quasi-particle, $p_{l}(t)$, can be represented by using the diffusive scaling function $\tilde{p}(x)$ as
\begin{eqnarray}
p_{l}(t) = t^{-1/2} \tilde{p}\Bigl((l-l_{0})t^{-1/2}\Bigr).
\end{eqnarray}
We show $\mathrm{ln}[t^{1/2}p_{l}(t)]$ against $(l-l_{0})t^{-1/2}$ in Fig.~\ref{fig:scaling}.
As shown in Fig.~\ref{fig:scaling}(a), for $A_{J} =0$, $\mathrm{ln}[t^{1/2}p_{l}(t)]$ have two symmetric peaks moving outward over time, and one peak located at the center.  
Although the symmetric peaks get smaller with time, they remain even in the long time scale. 
The shape of the central peak of $\mathrm{ln}[t^{1/2}p_{l}(t)]$ is time-independent and is a quadratic function of $(l-l_{0})t^{-1/2}$ in $|(l-l_{0})t^{-1/2}|\lesssim 3$.
This quadratic form verifies that the distribution around the central peak belongs to the same class as particle distributions for random walks.
For finite $A_{J}$, as shown in Figs.~\ref{fig:scaling}(b) and (c), the outer peaks in $\mathrm{ln}[t^{1/2}p_{l}(t)]$ become indistinct and almost disappear after a sufficiently long time.  
In addition, as seen in Fig.~\ref{fig:scaling}(c), the central peak in $\mathrm{ln}[t^{1/2}p_{l}(t)]$ deviates from its early form and converges to another diffusive scaling function.
It is verified that the quasi-particle becomes diffusive by the quasi-periodicity not only in the viewpoint of the dynamical exponent but also in the viewpoint of the probability distribution.


\subsection{Spin-Spin Correlation Function}
In the semi-classical framework introduced by Sachdev, correlation functions are estimated from the classical trajectories of quasi-particles.
To connect the quasi-particle dynamics analyzed above to the relaxation of correlations, we calculate a spin-spin correlation function by using Wick's theorem.
We consider the time dependence of a spin-spin correlation defined as 
\begin{eqnarray}
C_{l}(t) = \bra{x}\hat{\sigma}_{L/2-l/2}^{x}(t) \hat{\sigma}_{L/2+l/2}^{x}(t) \ket{x}.
\end{eqnarray}
At the initial time, no transverse field is applied to the system, and the state,  $\ket{x}$, is in one of the degenerate ground states, giving $C_{l}(0)=1$. 
At $t=0$, the transverse field is suddenly switched to a finite value with temporal fluctuation.
As described in Sec.~III.A, the quasi-particle propagates super-diffusively for certain time intervals in the short time scale, and its dynamics becomes to diffusive in the long time scale.
From the viewpoint of the semi-classical framework, the relaxations of the correlation functions are expected to have the same features as those of the quasi-particle dynamics.

The correlation function can be written in the Majorana representation as
\begin{eqnarray}
\bra{x}\hat{\sigma}_{i}^{x}(t) \hat{\sigma}_{j}^{x}(t) \ket{x} \propto \bra{x}\hat{\gamma}_{2i}(t)\hat{\gamma}_{2i+1}(t)\cdots \hat{\gamma}_{2j-1}(t)\ket{x}.\nonumber \\
\end{eqnarray}
Wick's theorem then enables us to calculate $C_{l}(t)$ by the Pfaffian $\mathrm{Pf}[\mathsf{X}(t)]$, where the skew matrix $\mathsf{X}(t)$ is defined by
\begin{eqnarray}
\mathsf{X}_{ij}(t) = \left\{
\begin{array}{ll}
0 & (i=j) \\
\bra{x} \hat{\gamma}_{i}(t)\hat{\gamma}_{j}(t)\ket{x}  & (i \neq j)
\end{array}
\right. .
\end{eqnarray}
\begin{figure}[t]
\centering
\includegraphics[keepaspectratio, scale=0.35]{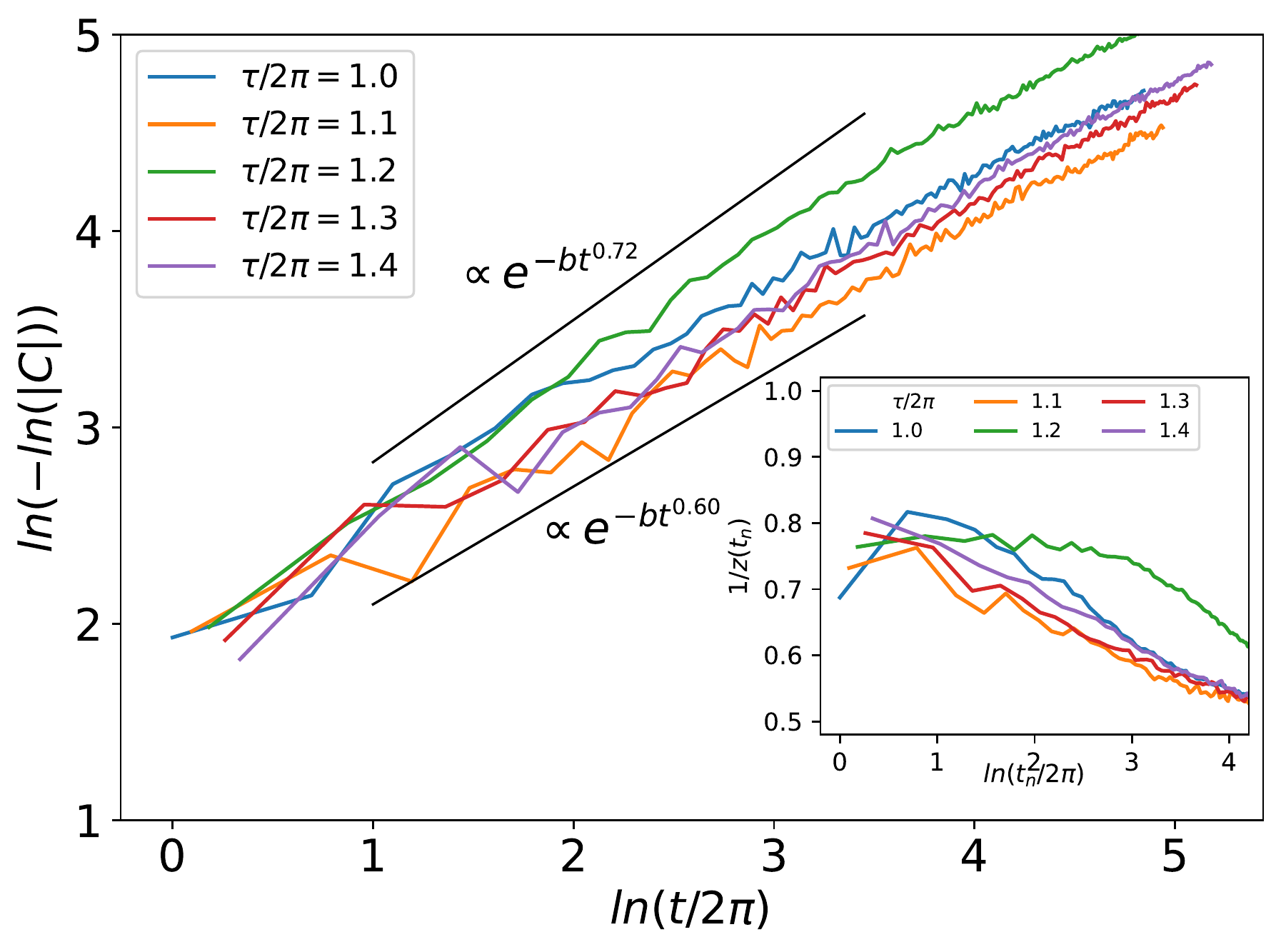}
\caption{Time dependence of the spin-spin correlation function with  $L=1024,~l=510,~J=1,~A_{J}=0.5,~h=1,~N_{\mathrm{samp}}=1000$ and $\tau/2\pi = 1.0-1.4$. The inset depicts the time-dependence of the dynamical exponents for the same setting as depicted in Fig.~2(a).
}
\label{fig:spin_correlation}
\end{figure}

Figure~\ref{fig:spin_correlation} shows numerical results for several time intervals $\tau$. 
Because the decay of the spin-spin correlation function over time is expected to be 
\begin{eqnarray}
C_{l}(t) \sim \mathrm{e}^{-bt^{a}},
\end{eqnarray}
we estimate the exponent $a$ from linear fitting by the least square method.
After a long time, the value of $a$ for $\tau/2\pi = 1.2$ approaches the same value as for the other $\tau$. 
The exponent $a=0.724$ for $\tau/2\pi = 1.2$, while $a\sim 0.6$ for the other $\tau$ in the short time scale.
These exponents $a$ are close to the dynamical exponents $1/z(t)$, which are depicted in the inset of Fig.~4, in the short time scale, $\mathrm{ln}(t/2\pi) \lesssim 3$.

\section{THEORY OF STROBOSCOPIC EIGENVECTORS}
\begin{figure}[t]
\centering
\includegraphics[keepaspectratio, scale=0.4]{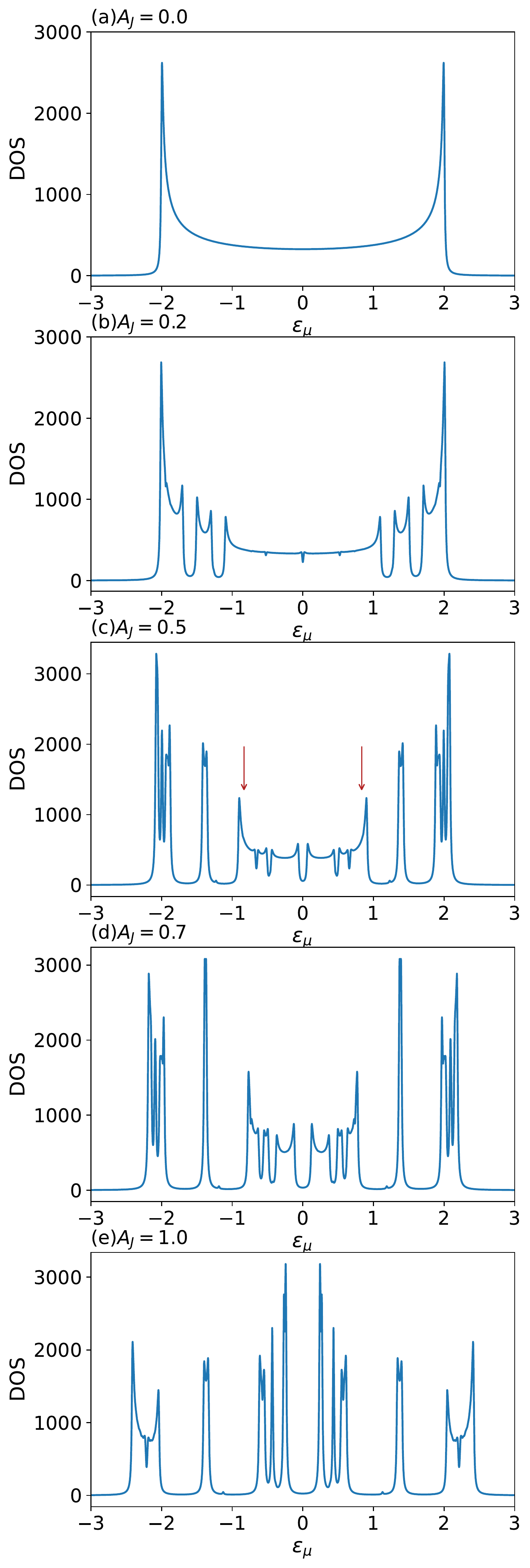}
\caption{
Density of states (DOS) of the matrix $\mathrm{H}$ with $L= 2048, h = 1, J= 1$, and $A_{J} =0, 0.2 , 0.5, 0.7, 1$.
The introduction of  $A_{J}$ to the TFIM ($A_{J} = 0$) causes gaps in the continuous density of states. 
The red arrows in the DOS for $A_{J}=0.5$ depict peaks corresponding to $\epsilon_{\mu}=\pm 0.833$.  
}
\label{fig:energy}
\end{figure}

\begin{figure}[h]
\centering
\includegraphics[keepaspectratio, scale=0.6]{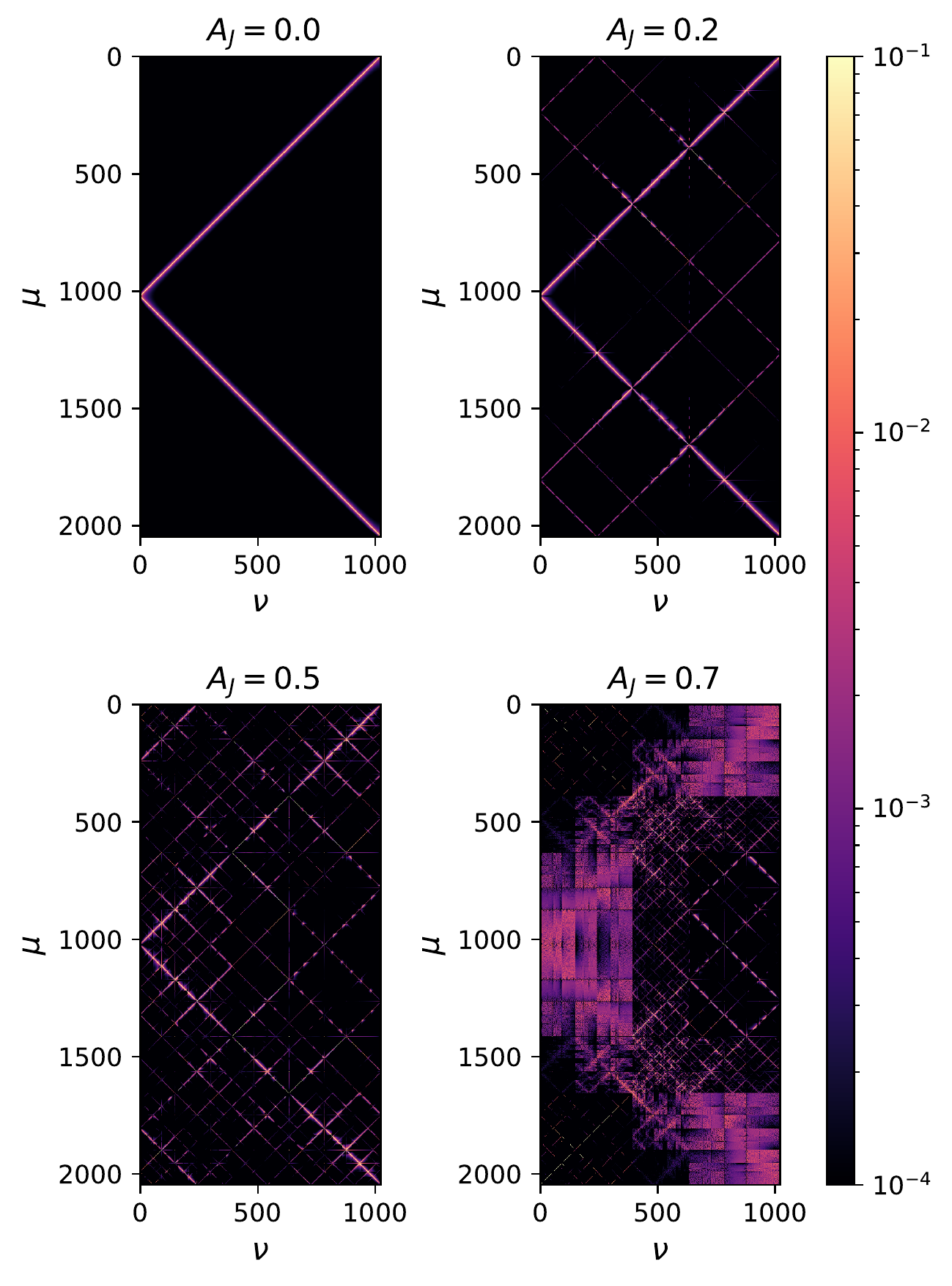}
\caption{
Overlap between the eigenvectors of $H_{+}$ and $H_{-}$. 
The vertical axis $\mu$ and horizontal axis $\nu$ represent the numbers of eigenvectors of $H_{+}$ and $H_{-}$, respectively.
}
\label{fig:occupancy}
\end{figure}

In this section, we review the theory of stroboscopic eigenvectors as adapted for the TFIM by Ro\'{o}sz \cite{Roosz}, and we extend the theory to the QP-TFIM. 
The key to understanding the above results is to grasp how an eigenvector of $\mathsf{H}_{\pm}$ can be written with eigenvectors of $\mathsf{H}_{\mp}$.
For the TFIM, the eigenvectors of $\mathsf{H}_{\pm}$ are spanned by those of $\mathsf{H}_{\mp}$ in a simple manner, which leads to the non-diffusive quasi-particle dynamics. 
For the QP-TFIM, the finite quasi-periodicity prevents such a simple manner, and this lack of the simple relation of overlap between the eigenvectors of $\mathsf{H}_{+}$ and $\mathsf{H}_{-}$ leads to diffusive behavior in the long time scale for any fluctuation time.

\subsection{Theory for TFIM}
The temporally random fields in question are considered by choosing either $\mathsf{O}_{+}$ or $\mathsf{O}_{-}$ at each duration as a time-evolution operator.
Generally, because an eigenvector of $\mathsf{O}_{+}$ is not an eigenvector of $\mathsf{O}_{-}$, after the operation of $\mathsf{O}_{-}$, an eigenvector is no longer an eigenvector of both $\mathsf{O}_{+}$ and $\mathsf{O}_{-}$.
After the random operations of $\mathsf{O}_{\pm}$, the system exhibits the diffusive dynamics.
In Ro\'{o}sz's theory, if one of the eigenvectors of $\mathsf{H}_{+}$ is an eigenvector of both $\mathsf{O}_{+}$ and $\mathsf{O}_{-}$, then it will survive even under random operations of $\mathsf{O}_{\pm}$.
We review this theory more precisely below.

In this section, $\mathsf{H}_{s}$ and $\mathsf{H}_{-s}$ denote the Hamiltonians, where $s$ represents the sign of $h(t)$.
For the TFIM with the periodic boundary condition, one can find that the eigenvectors of $\mathsf{H}_{s}$, and $\mathsf{H}_{-s}$ are related by 
\begin{eqnarray}
\boldsymbol{u}^{s}_{\mu}= c_{\nu\mu}\boldsymbol{u}^{-s}_{\nu}+c_{\overline{\nu}\mu}\boldsymbol{u}^{-s}_{\overline{\nu}} \label{eq:span_TFIM}.
\end{eqnarray}
Here, $\boldsymbol{u}^{s}_{\mu}, \boldsymbol{u}^{-s}_{\nu}$ are the respective eigenvectors of $\mathsf{H}_{s}, \mathsf{H}_{-s}$, and $c_{\nu\mu}=(\boldsymbol{u}^{-s}_{\nu}, \boldsymbol{u}^{s}_{\mu})$, where $(\cdot, \cdot)$ denotes an inner product as $(\boldsymbol{a},\boldsymbol{b}) = \sum_{n}a^{*}_{n}b_{n}$. 
The respective eigenvalues corresponding to $\boldsymbol{u}^{\pm s}_{\mu}$ are denoted by $\epsilon^{\pm s}_{\mu}$, and $\overline{\mu}$ represents the eigenvector with eigenvalue $\epsilon^{\pm s}_{\mu}=-\epsilon^{\pm s}_{\overline{\mu}}$.

The correlation functions given by Eq.~(\ref{eq:def_G}) can be written in linear combinations of the eigenvectors of $\mathsf{H}_{\pm s}$.
The time evolution is implemented by random operations of $\mathsf{O}_{s}$ or $\mathsf{O}_{-s}$ as in Eq.~(\ref{eq:random_operation}).
We now check whether $\boldsymbol{u}_{\mu}^{s}$ survives under the operation of both $\mathsf{O}_{s}$ and $\mathsf{O}_{-s}$.  
The operation of $\mathsf{O}_{s} = \mathrm{e}^{-\mathrm{i}\mathrm{H}_{s}\tau}$ only changes a coefficient and does not change the eigenvector itself.  
By contrast,  $\mathsf{O}_{-s}= \mathrm{e}^{-\mathrm{i}\mathrm{H}_{-s}\tau}$ operates on  $\boldsymbol{u}^{s}_{\mu}$ as
\begin{eqnarray}
\mathsf{O}_{-s}\boldsymbol{u}^{s}_{\mu}= \mathrm{e}^{-\mathrm{i}\epsilon^{-s}_{\nu}\tau}c_{\nu\mu}\boldsymbol{u}^{-s}_{\nu}+\mathrm{e}^{\mathrm{i}\epsilon^{-s}_{\nu}\tau}c_{\bar{\nu}\mu}\boldsymbol{u}^{-s}_{\bar{\nu}}.
\end{eqnarray}
By assuming the relation as
\begin{eqnarray}
\epsilon^{-s}_{\nu}\tau = m_{-s}\pi, ~m_{-s} = 1, 2,\cdots, \label{eq:assump}
\end{eqnarray} 
we obtain $\boldsymbol{u}^{s}_{\mu}$ as an eigenvector of the time-evolution operator $\mathsf{O}_{-s}$ even though $\boldsymbol{u}^{s}_{\mu}$, is not an eigenvector of $\mathsf{H}_{-s}$.

While the Hamiltonian of the TFIM has no eigenvector satisfying the assumption for $\tau < \pi/2$, the Hamiltonian does have eigenvectors whose eigenvalues satisfy the assumption for $\tau \geq \pi/2$, as depicted in Fig.~\ref{fig:energy}(a). 
For $\tau \geq \pi/2$, the correlation function includes the components of the eigenvectors satisfying the assumption for $\tau$, which are never disturbed by the temporal noise.
Such eigenvectors that survive under the temporal noise contribute to the non-diffusive dynamics of the quasi-particles for $\tau \geq \pi/2$.

\subsection{Theory for QP-TFIM}
For the QP-TFIM, we obtain the non-diffusive dynamics in the short time scale and the relaxation of the dynamical exponents to the diffusive regime in the long time scale.  
Here, we extend Ro\'{o}sz's theory to the QP-TFIM to explain the difference between the dynamics in the TFIM and the QP-TFIM.

In the presence of the quasi-periodicity, Eq.~(\ref{eq:span_TFIM}) no longer holds.
For the QP-TFIM, more than two pairs of the eigenvectors are necessary to describe the eigenvectors of the other Hamiltonian.
To check this expectation, we define overlap between the eigenvectors of $\mathsf{H}_{s}$ and $\mathsf{H}_{-s}$ as
\begin{eqnarray}
M_{\mu\nu}:= \bigl|(\boldsymbol{u}^{s}_{\mu}, \boldsymbol{u}^{-s}_{\nu})\bigr|^{2} + \bigl|(\boldsymbol{u}^{s}_{\mu},\boldsymbol{u}^{-s}_{\bar{\nu}})\bigr|^{2}.
\end{eqnarray}
Here, $M_{\mu\nu}$ represents how $\boldsymbol{u}^{s}_{\mu}$ is spanned by the pairs of $\boldsymbol{u}^{-s}_{\nu}$.
Figure~\ref{fig:occupancy} shows calculated results of $M_{\mu\nu}$.
In the TFIM case ($A_{J} = 0.0$), we find that a single eigenvector $\boldsymbol{u}^{A}_{\mu}$ has an amplitude only for a certain $\nu$, which shows that $\boldsymbol{u}^{s}_{\mu}$ is spanned only by the pair $\{\boldsymbol{u}_{\nu}^{-s},\boldsymbol{u}_{\overline{\nu}}^{-s}\}$.
A finite $A_{J}$ leads to fine structures in addition to the clear line of the TFIM, and such structures get smeared with increasing $A_{J}$.
This result shows that more than two pairs $\{\boldsymbol{u}_{\nu}^{-s},\boldsymbol{u}_{\overline{\nu}}^{-s}\}$ are necessary to span an eigenvector $\boldsymbol{u}^{s}_{\mu}$ in the QP-TFIM.

Consider the simplest case, in which $\boldsymbol{u}_{\mu}^{s}$ is spanned by only two pairs, $\{\boldsymbol{u}_{\nu}^{-s}, \boldsymbol{u}_{\overline{\nu}}^{-s}\}$ and $\{\boldsymbol{u}_{\lambda}^{-s}, \boldsymbol{u}_{\overline{\lambda}}^{-s}\}$:
\begin{eqnarray}
\boldsymbol{u}_{\mu}^{s} = c_{\nu\mu}\boldsymbol{u}_{\nu}^{-s} + c_{\overline{\nu}\mu}\boldsymbol{u}_{\overline{\nu}}^{-s}+ c_{\lambda\mu}\boldsymbol{u}_{\lambda}^{-s} + c_{\overline{\lambda}\mu}\boldsymbol{u}_{\overline{\lambda}}^{-s}.
\label{eq:span_QP-TFIM}
\end{eqnarray}
The Operation of $\mathsf{O}_{-s}$ on $\boldsymbol{u}_{\mu}^{s}$ leads to
\begin{eqnarray}
\mathsf{O}_{-s}\boldsymbol{u}_{\mu}^{s} 
&=& 
c_{\nu\mu}\mathrm{e}^{-\mathrm{i}\epsilon^{-s}_{\nu}\tau}\boldsymbol{u}_{\nu}^{-s} 
+ c_{\overline{\nu}\mu}\mathrm{e}^{\mathrm{i}\epsilon^{-s}_{\nu}\tau}\boldsymbol{u}_{\overline{\nu}}^{-s}\nonumber\\
&\quad&+~
 c_{\lambda\mu}\mathrm{e}^{-\mathrm{i}\epsilon^{-s}_{\lambda}\tau}\boldsymbol{u}_{\lambda}^{-s} 
+ c_{\overline{\lambda}\mu}\mathrm{e}^{\mathrm{i}\epsilon^{-s}_{\lambda}\tau}\boldsymbol{u}_{\overline{\lambda}}^{-s}.
\label{eq:operation_B}
\end{eqnarray}
Here, even if either $\mathrm{e}^{\pm\mathrm{i}\epsilon^{-s}_{\nu}\tau}$ or $\mathrm{e}^{\pm\mathrm{i}\epsilon^{-s}_{\lambda}\tau}$ equals $\pm 1$, the other one cannot be $\pm 1$.
In this study, we do not consider the case of $\epsilon^{-s}_{\lambda}=n_{}\epsilon^{-s}_{\nu},~n_{}\in\mathbb{Z}$.
Thus, when $\boldsymbol{u}^{s}_{\mu}$ is spanned by multiple pairs $\{\boldsymbol{u}_{\nu}^{-s}, \boldsymbol{u}_{\overline{\nu}}^{-s}\}$, $\boldsymbol{u}^{s}_{\mu}$ cannot be an eigenvector of $\mathsf{O}_{-s}$.

The quasi-particle of the TFIM avoids the diffusive dynamics even in the long time scale.
The reason is that the eigenvector of the TFIM is preserved under the random operations of $\mathsf{O}_{\pm}$ because of the simple overlap shown in Fig.~\ref{fig:occupancy}.
However, the QP-TFIM has more complex overlap than the TFIM, and there is no preservation of the eigenvector under the random operation of $\mathsf{O}_{\pm}$.  
Thus, the quasi-particle dynamics of the QP-TFIM relaxes to the diffusive regime in the long time scale.

The relation given by Eqs.~(\ref{eq:span_TFIM}) and (\ref{eq:span_QP-TFIM}) determines the speed of the relaxation shown in Fig.~\ref{fig:zeff}.
For small $A_{J}$, the overlap in Fig.~\ref{fig:occupancy} shows clear lines, which corresponds to the dominance of the first and second terms in Eq.~(\ref{eq:span_QP-TFIM}), where as the other terms are tiny.
Because $\boldsymbol{u}_{\mu}^{s}$ is almost the same as the eigenvector of $\mathsf{O}_{-s}$ in the case of small $A_{J}$, the single operation of $\mathsf{O}_{-s}$ transforms $\boldsymbol{u}_{\mu}^{s}$ to a vector that is slightly different from $\boldsymbol{u}_{\mu}^{s}$. 
Such slight changes accumulate through the random operations of $\mathsf{O}_{\pm}$, and the eigenvector loses its quantum nature in the long time scale.
On the other hand, for large $A_{J}$, the third, fourth, and subsequent terms are comparable to the first and second terms, and even a single operation of $\mathsf{O}_{-s}$ transforms $\boldsymbol{u}_{\mu}^{s}$ to a vector that is greatly different from $\boldsymbol{u}_{\mu}^{s}$.
The eigenvector also reaches a mixed state in the long time scale, where the speed to reach the mixed state is faster than for small $A_{J}$.  

Note that we also obtain the slower relaxation of the dynamical exponent for $\tau/2\pi=1.2$ and $2.4$ than for the other $\tau$.
The slower relaxation can be explained by considering the modification in eigenenergy structures with increasing $A_{J}$ in Fig.~\ref{fig:energy}.
The corresponding energies with $\tau/2\pi=1.2$ and $2.4$ are $\epsilon_{-s} = 0.833$ for $m_{-s} = 2$ and $m_{-s}=4$, respectively.
Small peaks at $\epsilon_{-s} = \pm 0.833$ can be seen in Fig.~\ref{fig:energy}.
Thus, the assumption is satisfied for $\tau/2\pi=1.2$ and $2.4$, but not for the other $\tau$; accordingly the super-diffusive dynamics occurs only for $\tau/2\pi=1.2$ and $2.4$.
  
\section{summary}
In this paper, we have studied the quasi-particle dynamics in a quasi-periodic Ising model with temporally random fields (QP-TFIM).
Specifically, we analyzed the stroboscopic time evolution of the probability distributions of the quasi-particles after the quenches.
Our results can be summarized as follows.
The quasi-particles exhibit the two different behaviors depending on the time scale.
The short-time behavior depends on the time interval, and the dynamical exponents are higher  for the certain intervals than for the other intervals.  
In contrast, the long-time behavior becomes diffusive for any time interval.
We showed the time dependence of the spin-spin correlation function.
The exponents for the relaxation of the correlation function are consistent with the dynamical exponents of the quasi-particles. 
These results can be explained by the overlap of the eigenvectors of each Hamiltonian. 
For the TFIM, the structure of the overlap is simple.
For the QP-TFIM, however, the quasi-periodicity complicates the overlapping structure, which causes the diffusive dynamics of quasi-particles in the long time scale.

As mentioned in the introduction, both quasi-periodicity and random disorder are origins of localization, but the energy levels in these systems have statistical properties different from each other.
The energy levels in quasi-periodic systems are uniquely determined for a single set of Hamiltonian's parameters, which lead the slow relaxation of the quasi-particle dynamics as discussed in this paper.
On the other hand, the energy levels in random disordered systems are not fixed for certain parameters of the Hamiltonian but depend on configurations of these random potentials. 
For this reason, we expect that certain intervals leading to the slow decay of the dynamical exponent do not exist in random disordered systems.
In this study, we have treated the randomly fluctuating fields with certain intervals, whose situation is related to that under periodic drivings, as in Floquet systems \cite{Floquet_review}.
Recently, the Floquet systems are actively studied in the context of the thermalization \cite{D'Alessio, Ponte, Zhang_Floquet, Regnault, Luitz_Floquet}.
It is worthwhile investigating the quasi-particle dynamics for the Foquet systems and the overlap relation between eigenvectors of the time evolution operators. 
These analyses are left for future works.

\begin{acknowledgments}
We thank Lev Vidmar for useful discussions with us.
K.~O. is supported by  GP-Spin at Tohoku University and JST SPRING, Grant Number JPMJSP2114.
Y.~M. is supported by JPSJ KAKENHI (Nos. 19K14662, 22H01221).
H.~M. is supported by JPSJ KAKENHI (Nos. 21K03380, 21H04446, 21H03455, 20K03769) and CSIS at Tohoku University.
\end{acknowledgments}

\nocite{*}

\bibliography{apssamp}

\providecommand{\noopsort}[1]{}\providecommand{\singleletter}[1]{#1}%
\begin{thebibliography}{49}%
\makeatletter
\providecommand \@ifxundefined [1]{%
 \@ifx{#1\undefined}
}%
\providecommand \@ifnum [1]{%
 \ifnum #1\expandafter \@firstoftwo
 \else \expandafter \@secondoftwo
 \fi
}%
\providecommand \@ifx [1]{%
 \ifx #1\expandafter \@firstoftwo
 \else \expandafter \@secondoftwo
 \fi
}%
\providecommand \natexlab [1]{#1}%
\providecommand \enquote  [1]{``#1''}%
\providecommand \bibnamefont  [1]{#1}%
\providecommand \bibfnamefont [1]{#1}%
\providecommand \citenamefont [1]{#1}%
\providecommand \href@noop [0]{\@secondoftwo}%
\providecommand \href [0]{\begingroup \@sanitize@url \@href}%
\providecommand \@href[1]{\@@startlink{#1}\@@href}%
\providecommand \@@href[1]{\endgroup#1\@@endlink}%
\providecommand \@sanitize@url [0]{\catcode `\\12\catcode `\$12\catcode
  `\&12\catcode `\#12\catcode `\^12\catcode `\_12\catcode `\%12\relax}%
\providecommand \@@startlink[1]{}%
\providecommand \@@endlink[0]{}%
\providecommand \url  [0]{\begingroup\@sanitize@url \@url }%
\providecommand \@url [1]{\endgroup\@href {#1}{\urlprefix }}%
\providecommand \urlprefix  [0]{URL }%
\providecommand \Eprint [0]{\href }%
\providecommand \doibase [0]{https://doi.org/}%
\providecommand \selectlanguage [0]{\@gobble}%
\providecommand \bibinfo  [0]{\@secondoftwo}%
\providecommand \bibfield  [0]{\@secondoftwo}%
\providecommand \translation [1]{[#1]}%
\providecommand \BibitemOpen [0]{}%
\providecommand \bibitemStop [0]{}%
\providecommand \bibitemNoStop [0]{.\EOS\space}%
\providecommand \EOS [0]{\spacefactor3000\relax}%
\providecommand \BibitemShut  [1]{\csname bibitem#1\endcsname}%
\let\auto@bib@innerbib\@empty
\bibitem [{\citenamefont {Anderson}(1958)}]{Anderson}%
  \BibitemOpen
  \bibfield  {author} {\bibinfo {author} {\bibfnamefont {P.~W.}\ \bibnamefont
  {Anderson}},\ }\href@noop {} {\bibfield  {journal} {\bibinfo  {journal}
  {Phys.\ Rev.}\ }\textbf {\bibinfo {volume} {109}},\ \bibinfo {pages} {1492}
  (\bibinfo {year} {1958})}\BibitemShut {NoStop}%
\bibitem [{\citenamefont {{E. Abrahams, P. W. Anderson, D. C. Licciardello, and
  T. V. Ramakrishnan}}(1979)}]{Abrahams}%
  \BibitemOpen
  \bibfield  {author} {\bibinfo {author} {\bibnamefont {{E. Abrahams, P. W.
  Anderson, D. C. Licciardello, and T. V. Ramakrishnan}}},\ }\href@noop {}
  {\bibfield  {journal} {\bibinfo  {journal} {Phys.\ Rev.\ Lett.}\ }\textbf
  {\bibinfo {volume} {42}},\ \bibinfo {pages} {673} (\bibinfo {year}
  {1979})}\BibitemShut {NoStop}%
\bibitem [{\citenamefont {{F. Igl\'{o}i and H.
  Rieger}}(1998)}]{Igloi_randomTFIM}%
  \BibitemOpen
  \bibfield  {author} {\bibinfo {author} {\bibnamefont {{F. Igl\'{o}i and H.
  Rieger}}},\ }\href@noop {} {\bibfield  {journal} {\bibinfo  {journal} {Phys.\
  Rev.\ B}\ }\textbf {\bibinfo {volume} {57}},\ \bibinfo {pages} {11404}
  (\bibinfo {year} {1998})}\BibitemShut {NoStop}%
\bibitem [{\citenamefont {{F. Evers, and A. D. Mirlin}}(2008)}]{Evers}%
  \BibitemOpen
  \bibfield  {author} {\bibinfo {author} {\bibnamefont {{F. Evers, and A. D.
  Mirlin}}},\ }\href@noop {} {\bibfield  {journal} {\bibinfo  {journal} {Rev.\
  Mod.\ Phys.}\ }\textbf {\bibinfo {volume} {80}},\ \bibinfo {pages} {1355}
  (\bibinfo {year} {2008})}\BibitemShut {NoStop}%
\bibitem [{\citenamefont {{R. Nandkishore, and D. A.
  Huse}}(2015)}]{Nandkishore}%
  \BibitemOpen
  \bibfield  {author} {\bibinfo {author} {\bibnamefont {{R. Nandkishore, and D.
  A. Huse}}},\ }\href@noop {} {\bibfield  {journal} {\bibinfo  {journal} {Annu.
  \ Rev. \ Condens. \ Matter \ Phys.}\ }\textbf {\bibinfo {volume} {6}},\
  \bibinfo {pages} {15} (\bibinfo {year} {2015})}\BibitemShut {NoStop}%
\bibitem [{\citenamefont {{F. Alet, and N. Laflorencie}}(2018)}]{Alet}%
  \BibitemOpen
  \bibfield  {author} {\bibinfo {author} {\bibnamefont {{F. Alet, and N.
  Laflorencie}}},\ }\href@noop {} {\bibfield  {journal} {\bibinfo  {journal}
  {C. \ R. \ Phys.}\ }\textbf {\bibinfo {volume} {19}},\ \bibinfo {pages} {498}
  (\bibinfo {year} {2018})}\BibitemShut {NoStop}%
\bibitem [{\citenamefont {{D. A. Abanin, E. Altman, I. Bloch, and M.
  Serbyn}}(2019)}]{Abanin}%
  \BibitemOpen
  \bibfield  {author} {\bibinfo {author} {\bibnamefont {{D. A. Abanin, E.
  Altman, I. Bloch, and M. Serbyn}}},\ }\href@noop {} {\bibfield  {journal}
  {\bibinfo  {journal} {Rev. Mod. Phys.}\ }\textbf {\bibinfo {volume} {91}},\
  \bibinfo {pages} {021001} (\bibinfo {year} {2019})}\BibitemShut {NoStop}%
\bibitem [{\citenamefont {{Y. Aharonov, L. Davidovich, and N.
  Zagury}}(1993)}]{Aharonov}%
  \BibitemOpen
  \bibfield  {author} {\bibinfo {author} {\bibnamefont {{Y. Aharonov, L.
  Davidovich, and N. Zagury}}},\ }\href@noop {} {\bibfield  {journal} {\bibinfo
   {journal} {Phys.\ Rev.\ A}\ }\textbf {\bibinfo {volume} {48}},\ \bibinfo
  {pages} {1687} (\bibinfo {year} {1993})}\BibitemShut {NoStop}%
\bibitem [{\citenamefont {{A. Joye, and M. Merkli}}(2010)}]{Joye}%
  \BibitemOpen
  \bibfield  {author} {\bibinfo {author} {\bibnamefont {{A. Joye, and M.
  Merkli}}},\ }\href@noop {} {\bibfield  {journal} {\bibinfo  {journal} {J. \
  Stat. \ Phys.}\ }\textbf {\bibinfo {volume} {140}},\ \bibinfo {pages} {1}
  (\bibinfo {year} {2010})}\BibitemShut {NoStop}%
\bibitem [{\citenamefont {{N. Konno}}(2010)}]{Konno_1}%
  \BibitemOpen
  \bibfield  {author} {\bibinfo {author} {\bibnamefont {{N. Konno}}},\
  }\href@noop {} {\bibfield  {journal} {\bibinfo  {journal} {Quantum \ Inf. \
  Process.}\ }\textbf {\bibinfo {volume} {9}},\ \bibinfo {pages} {387}
  (\bibinfo {year} {2010})}\BibitemShut {NoStop}%
\bibitem [{\citenamefont {{N. Konno}}(2009)}]{Konno_2}%
  \BibitemOpen
  \bibfield  {author} {\bibinfo {author} {\bibnamefont {{N. Konno}}},\
  }\href@noop {} {\bibfield  {journal} {\bibinfo  {journal} {Quantum \ Inf. \
  Process.}\ }\textbf {\bibinfo {volume} {8}},\ \bibinfo {pages} {387}
  (\bibinfo {year} {2009})}\BibitemShut {NoStop}%
\bibitem [{\citenamefont {{H. Obuse, and N. Kawakami}}(2011)}]{Obuse}%
  \BibitemOpen
  \bibfield  {author} {\bibinfo {author} {\bibnamefont {{H. Obuse, and N.
  Kawakami}}},\ }\href@noop {} {\bibfield  {journal} {\bibinfo  {journal}
  {Phys. \ Rev. \ B}\ }\textbf {\bibinfo {volume} {84}},\ \bibinfo {pages}
  {195139} (\bibinfo {year} {2011})}\BibitemShut {NoStop}%
\bibitem [{\citenamefont {{A. Ahlbrecht, V. B. Scholz, and A. H.
  Werner}}(2011)}]{Ahlbrecht_spatio_1}%
  \BibitemOpen
  \bibfield  {author} {\bibinfo {author} {\bibnamefont {{A. Ahlbrecht, V. B.
  Scholz, and A. H. Werner}}},\ }\href@noop {} {\bibfield  {journal} {\bibinfo
  {journal} {J. Math Phys.}\ }\textbf {\bibinfo {volume} {52}},\ \bibinfo
  {pages} {102201} (\bibinfo {year} {2011})}\BibitemShut {NoStop}%
\bibitem [{\citenamefont {{G.Abal, R.Donangelo, F.Severo, and
  R.Siri}}(2008)}]{Abal}%
  \BibitemOpen
  \bibfield  {author} {\bibinfo {author} {\bibnamefont {{G.Abal, R.Donangelo,
  F.Severo, and R.Siri}}},\ }\href@noop {} {\bibfield  {journal} {\bibinfo
  {journal} {Phys. \ A \ Stat. \ Mech. \ App.}\ }\textbf {\bibinfo {volume}
  {387}},\ \bibinfo {pages} {335} (\bibinfo {year} {2008})}\BibitemShut
  {NoStop}%
\bibitem [{\citenamefont {{C. M. Chandrashekar, R. Srikanth, and S.
  Banerjee}}(2007)}]{Chandrashekar}%
  \BibitemOpen
  \bibfield  {author} {\bibinfo {author} {\bibnamefont {{C. M. Chandrashekar,
  R. Srikanth, and S. Banerjee}}},\ }\href@noop {} {\bibfield  {journal}
  {\bibinfo  {journal} {Phys. \ Rev. \ A}\ }\textbf {\bibinfo {volume} {76}},\
  \bibinfo {pages} {022316} (\bibinfo {year} {2007})}\BibitemShut {NoStop}%
\bibitem [{\citenamefont {{J. Ko\v{s}\'{i}k, V. Bu\v{z}ek, and M.
  Hillery}}(2006)}]{Kosik}%
  \BibitemOpen
  \bibfield  {author} {\bibinfo {author} {\bibnamefont {{J. Ko\v{s}\'{i}k, V.
  Bu\v{z}ek, and M. Hillery}}},\ }\href@noop {} {\bibfield  {journal} {\bibinfo
   {journal} {Phys. \ Rev. \ A}\ }\textbf {\bibinfo {volume} {74}},\ \bibinfo
  {pages} {022310} (\bibinfo {year} {2006})}\BibitemShut {NoStop}%
\bibitem [{\citenamefont {{D. Shapira, O. Biham, A. J. Bracken, and M.
  Hackett}}(2003)}]{Shapira}%
  \BibitemOpen
  \bibfield  {author} {\bibinfo {author} {\bibnamefont {{D. Shapira, O. Biham,
  A. J. Bracken, and M. Hackett}}},\ }\href@noop {} {\bibfield  {journal}
  {\bibinfo  {journal} {Phys. \ Rev. \ A}\ }\textbf {\bibinfo {volume} {68}},\
  \bibinfo {pages} {062315} (\bibinfo {year} {2003})}\BibitemShut {NoStop}%
\bibitem [{\citenamefont {{A. Romanelli, R. Siri, G. Abal, A. Auyuanet, and R.
  Donangelo}}(2005)}]{Romanelli}%
  \BibitemOpen
  \bibfield  {author} {\bibinfo {author} {\bibnamefont {{A. Romanelli, R. Siri,
  G. Abal, A. Auyuanet, and R. Donangelo}}},\ }\href@noop {} {\bibfield
  {journal} {\bibinfo  {journal} {Phys. \ A \ Stat. \ Mech. \ App.}\ }\textbf
  {\bibinfo {volume} {347}},\ \bibinfo {pages} {137} (\bibinfo {year}
  {2005})}\BibitemShut {NoStop}%
\bibitem [{\citenamefont {{G. Leung, P. Knott, J. Bailey and V.
  Kendon}}(2010)}]{Leung}%
  \BibitemOpen
  \bibfield  {author} {\bibinfo {author} {\bibnamefont {{G. Leung, P. Knott, J.
  Bailey and V. Kendon}}},\ }\href@noop {} {\bibfield  {journal} {\bibinfo
  {journal} {New. \ J. \ Phys.}\ }\textbf {\bibinfo {volume} {12}},\ \bibinfo
  {pages} {1} (\bibinfo {year} {2010})}\BibitemShut {NoStop}%
\bibitem [{\citenamefont {{A. Ahlbrecht, C. Cedzich, R. Matjeschk, V. B.
  Scholz, A. H. Werner, and R. F. Werner}}(2012)}]{Ahlbrecht_temopral}%
  \BibitemOpen
  \bibfield  {author} {\bibinfo {author} {\bibnamefont {{A. Ahlbrecht, C.
  Cedzich, R. Matjeschk, V. B. Scholz, A. H. Werner, and R. F. Werner}}},\
  }\href@noop {} {\bibfield  {journal} {\bibinfo  {journal} {Quantum \ Inf. \
  Process.}\ }\textbf {\bibinfo {volume} {11}},\ \bibinfo {pages} {1219}
  (\bibinfo {year} {2012})}\BibitemShut {NoStop}%
\bibitem [{\citenamefont {{M. Montero}}(2016)}]{Montero}%
  \BibitemOpen
  \bibfield  {author} {\bibinfo {author} {\bibnamefont {{M. Montero}}},\
  }\href@noop {} {\bibfield  {journal} {\bibinfo  {journal} {Phys. \ Rev. \ A}\
  }\textbf {\bibinfo {volume} {93}},\ \bibinfo {pages} {062316} (\bibinfo
  {year} {2016})}\BibitemShut {NoStop}%
\bibitem [{\citenamefont {{C. K. Burrell, J. Eisert, and T. J.
  Osborne}}(2009)}]{Burrell}%
  \BibitemOpen
  \bibfield  {author} {\bibinfo {author} {\bibnamefont {{C. K. Burrell, J.
  Eisert, and T. J. Osborne}}},\ }\href@noop {} {\bibfield  {journal} {\bibinfo
   {journal} {Phys. \ Rev. \ A}\ }\textbf {\bibinfo {volume} {80}},\ \bibinfo
  {pages} {052319} (\bibinfo {year} {2009})}\BibitemShut {NoStop}%
\bibitem [{\citenamefont {{O. M\"{u}lken, and A. Blumen}}(2011)}]{Mulken}%
  \BibitemOpen
  \bibfield  {author} {\bibinfo {author} {\bibnamefont {{O. M\"{u}lken, and A.
  Blumen}}},\ }\href@noop {} {\bibfield  {journal} {\bibinfo  {journal} {Phys.
  \ Rep.}\ }\textbf {\bibinfo {volume} {502}},\ \bibinfo {pages} {37} (\bibinfo
  {year} {2011})}\BibitemShut {NoStop}%
\bibitem [{\citenamefont {{S. Sachdev, and A. P. Young}}(1997)}]{Sachdev}%
  \BibitemOpen
  \bibfield  {author} {\bibinfo {author} {\bibnamefont {{S. Sachdev, and A. P.
  Young}}},\ }\href@noop {} {\bibfield  {journal} {\bibinfo  {journal} {Phys. \
  Rev. \ Lett.}\ }\textbf {\bibinfo {volume} {78}},\ \bibinfo {pages} {2220}
  (\bibinfo {year} {1997})}\BibitemShut {NoStop}%
\bibitem [{\citenamefont {{H. Rieger, and F. Igl\'{o}i}}(2011)}]{Rieger}%
  \BibitemOpen
  \bibfield  {author} {\bibinfo {author} {\bibnamefont {{H. Rieger, and F.
  Igl\'{o}i}}},\ }\href@noop {} {\bibfield  {journal} {\bibinfo  {journal}
  {Phys. \ Rev. \ B}\ }\textbf {\bibinfo {volume} {84}},\ \bibinfo {pages}
  {165117} (\bibinfo {year} {2011})}\BibitemShut {NoStop}%
\bibitem [{\citenamefont {{F. Igl\'{o}i, G. Ro\'{o}sz, and Y.
  Lin}}(2013)}]{Igloi_quasicrystal}%
  \BibitemOpen
  \bibfield  {author} {\bibinfo {author} {\bibnamefont {{F. Igl\'{o}i, G.
  Ro\'{o}sz, and Y. Lin}}},\ }\href@noop {} {\bibfield  {journal} {\bibinfo
  {journal} {New \ J. \ Phys.}\ }\textbf {\bibinfo {volume} {15}},\ \bibinfo
  {pages} {023036} (\bibinfo {year} {2013})}\BibitemShut {NoStop}%
\bibitem [{\citenamefont {{G. Ro\'{o}sz, R. Juh\'{a}sz, and F.
  Igl\'{o}i}}(2016)}]{Roosz}%
  \BibitemOpen
  \bibfield  {author} {\bibinfo {author} {\bibnamefont {{G. Ro\'{o}sz, R.
  Juh\'{a}sz, and F. Igl\'{o}i}}},\ }\href@noop {} {\bibfield  {journal}
  {\bibinfo  {journal} {Phys. \ Rev. \ B}\ }\textbf {\bibinfo {volume} {93}},\
  \bibinfo {pages} {134305} (\bibinfo {year} {2016})}\BibitemShut {NoStop}%
\bibitem [{\citenamefont {{M. Ya. Azbel}}(1979)}]{Abzel}%
  \BibitemOpen
  \bibfield  {author} {\bibinfo {author} {\bibnamefont {{M. Ya. Azbel}}},\
  }\href@noop {} {\bibfield  {journal} {\bibinfo  {journal} {Phys.\ Rev.\
  Lett.}\ }\textbf {\bibinfo {volume} {43}},\ \bibinfo {pages} {1954} (\bibinfo
  {year} {1979})}\BibitemShut {NoStop}%
\bibitem [{\citenamefont {{S. Aubry, and G. Andr\'{e} }}(1980)}]{Aubry_Andre}%
  \BibitemOpen
  \bibfield  {author} {\bibinfo {author} {\bibnamefont {{S. Aubry, and G.
  Andr\'{e} }}},\ }\href@noop {} {\bibfield  {journal} {\bibinfo  {journal}
  {Ann.\ Isr. \ Phys.\ Soc.}\ }\textbf {\bibinfo {volume} {3}},\ \bibinfo
  {pages} {133} (\bibinfo {year} {1980})}\BibitemShut {NoStop}%
\bibitem [{\citenamefont {{S. Ostlund, R. Pandit, D. Rand, H. J. Schellnhuber,
  and E. D. Siggia}}(1983)}]{Ostlund}%
  \BibitemOpen
  \bibfield  {author} {\bibinfo {author} {\bibnamefont {{S. Ostlund, R. Pandit,
  D. Rand, H. J. Schellnhuber, and E. D. Siggia}}},\ }\href@noop {} {\bibfield
  {journal} {\bibinfo  {journal} {Phys.\ Rev.\ Lett.}\ }\textbf {\bibinfo
  {volume} {50}},\ \bibinfo {pages} {1873} (\bibinfo {year}
  {1983})}\BibitemShut {NoStop}%
\bibitem [{\citenamefont {{D. R. Hofstadter }}(1976)}]{Hofstadter}%
  \BibitemOpen
  \bibfield  {author} {\bibinfo {author} {\bibnamefont {{D. R. Hofstadter }}},\
  }\href@noop {} {\bibfield  {journal} {\bibinfo  {journal} {Phys.\ Rev. \ B}\
  }\textbf {\bibinfo {volume} {14}},\ \bibinfo {pages} {2239} (\bibinfo {year}
  {1976})}\BibitemShut {NoStop}%
\bibitem [{\citenamefont {{R. Ketzmerick, K. Kruse, F. Steinbach, and T.
  Geisel}}(1998)}]{Ketzmerick}%
  \BibitemOpen
  \bibfield  {author} {\bibinfo {author} {\bibnamefont {{R. Ketzmerick, K.
  Kruse, F. Steinbach, and T. Geisel}}},\ }\href@noop {} {\bibfield  {journal}
  {\bibinfo  {journal} {Phys.\ Rev. \ B}\ }\textbf {\bibinfo {volume} {58}},\
  \bibinfo {pages} {9881} (\bibinfo {year} {1998})}\BibitemShut {NoStop}%
\bibitem [{\citenamefont {{H. Hiramoto, and S. Abe}}(1988)}]{Hiramoto}%
  \BibitemOpen
  \bibfield  {author} {\bibinfo {author} {\bibnamefont {{H. Hiramoto, and S.
  Abe}}},\ }\href@noop {} {\bibfield  {journal} {\bibinfo  {journal} {J. \
  Phys. \ Soc. \ Jpn.}\ }\textbf {\bibinfo {volume} {57}},\ \bibinfo {pages}
  {230} (\bibinfo {year} {1988})}\BibitemShut {NoStop}%
\bibitem [{\citenamefont {{F. Pi\'{e}chon}}(1996)}]{Piechon}%
  \BibitemOpen
  \bibfield  {author} {\bibinfo {author} {\bibnamefont {{F. Pi\'{e}chon}}},\
  }\href@noop {} {\bibfield  {journal} {\bibinfo  {journal} {Phys. \ Rev. \
  Lett.}\ }\textbf {\bibinfo {volume} {76}},\ \bibinfo {pages} {4372} (\bibinfo
  {year} {1996})}\BibitemShut {NoStop}%
\bibitem [{\citenamefont {{F. Setiawan, D. L. Deng, and J. H.
  Pixley}}(2017)}]{Setiawan}%
  \BibitemOpen
  \bibfield  {author} {\bibinfo {author} {\bibnamefont {{F. Setiawan, D. L.
  Deng, and J. H. Pixley}}},\ }\href@noop {} {\bibfield  {journal} {\bibinfo
  {journal} {Phys.\ Rev.\ B}\ }\textbf {\bibinfo {volume} {96}},\ \bibinfo
  {pages} {104205} (\bibinfo {year} {2017})}\BibitemShut {NoStop}%
\bibitem [{\citenamefont {{G. Ro\'{o}sz, U. Divakaran, H. Rieger, and F.
  Igl\'{o}i}}(2014)}]{Roosz_quasiperiod}%
  \BibitemOpen
  \bibfield  {author} {\bibinfo {author} {\bibnamefont {{G. Ro\'{o}sz, U.
  Divakaran, H. Rieger, and F. Igl\'{o}i}}},\ }\href@noop {} {\bibfield
  {journal} {\bibinfo  {journal} {Phys.\ Rev.\ B}\ }\textbf {\bibinfo {volume}
  {90}},\ \bibinfo {pages} {184202} (\bibinfo {year} {2014})}\BibitemShut
  {NoStop}%
\bibitem [{\citenamefont {{S. Ganeshan, J. H. Pixley, and S.
  DasSarma}}(2015)}]{Ganeshan}%
  \BibitemOpen
  \bibfield  {author} {\bibinfo {author} {\bibnamefont {{S. Ganeshan, J. H.
  Pixley, and S. DasSarma}}},\ }\href@noop {} {\bibfield  {journal} {\bibinfo
  {journal} {Phys.\ Rev.\ Lett.}\ }\textbf {\bibinfo {volume} {114}},\ \bibinfo
  {pages} {146601} (\bibinfo {year} {2015})}\BibitemShut {NoStop}%
\bibitem [{\citenamefont {{J. Biddle and S. DasSarma}}(2010)}]{Biddle_pow_hop}%
  \BibitemOpen
  \bibfield  {author} {\bibinfo {author} {\bibnamefont {{J. Biddle and S.
  DasSarma}}},\ }\href@noop {} {\bibfield  {journal} {\bibinfo  {journal}
  {Phys.\ Rev.\ Lett.}\ }\textbf {\bibinfo {volume} {104}},\ \bibinfo {pages}
  {070601} (\bibinfo {year} {2010})}\BibitemShut {NoStop}%
\bibitem [{\citenamefont {{J. Biddle, D. J. Priour, Jr., B. Wang, and S.
  DasSarma}}(2011)}]{Biddle}%
  \BibitemOpen
  \bibfield  {author} {\bibinfo {author} {\bibnamefont {{J. Biddle, D. J.
  Priour, Jr., B. Wang, and S. DasSarma}}},\ }\href@noop {} {\bibfield
  {journal} {\bibinfo  {journal} {Phys.\ Rev.\ B}\ }\textbf {\bibinfo {volume}
  {83}},\ \bibinfo {pages} {075105} (\bibinfo {year} {2011})}\BibitemShut
  {NoStop}%
\bibitem [{\citenamefont {{S. Iyer, V. Oganesyan, G. Refael, and D. A.
  Huse}}(2013)}]{Iyer}%
  \BibitemOpen
  \bibfield  {author} {\bibinfo {author} {\bibnamefont {{S. Iyer, V. Oganesyan,
  G. Refael, and D. A. Huse}}},\ }\href@noop {} {\bibfield  {journal} {\bibinfo
   {journal} {Phys.\ Rev.\ B}\ }\textbf {\bibinfo {volume} {87}},\ \bibinfo
  {pages} {134202} (\bibinfo {year} {2013})}\BibitemShut {NoStop}%
\bibitem [{\citenamefont {{A. Chandran, and C. R. Laumann}}(2017)}]{Chandran}%
  \BibitemOpen
  \bibfield  {author} {\bibinfo {author} {\bibnamefont {{A. Chandran, and C. R.
  Laumann}}},\ }\href@noop {} {\bibfield  {journal} {\bibinfo  {journal}
  {Phys.\ Rev.\ X}\ }\textbf {\bibinfo {volume} {7}},\ \bibinfo {pages}
  {031061} (\bibinfo {year} {2017})}\BibitemShut {NoStop}%
\bibitem [{\citenamefont {{P. J. D. Crowley, A. Chandran, and C. R.
  Laumann}}(2018)}]{Crowley}%
  \BibitemOpen
  \bibfield  {author} {\bibinfo {author} {\bibnamefont {{P. J. D. Crowley, A.
  Chandran, and C. R. Laumann}}},\ }\href@noop {} {\bibfield  {journal}
  {\bibinfo  {journal} {Phys.\ Rev.\ Lett.}\ }\textbf {\bibinfo {volume}
  {120}},\ \bibinfo {pages} {175702} (\bibinfo {year} {2018})}\BibitemShut
  {NoStop}%
\bibitem [{\citenamefont {{U. Divakaran}}(2018)}]{Divakaran}%
  \BibitemOpen
  \bibfield  {author} {\bibinfo {author} {\bibnamefont {{U. Divakaran}}},\
  }\href@noop {} {\bibfield  {journal} {\bibinfo  {journal} {Phys.\ Rev.\ E}\
  }\textbf {\bibinfo {volume} {98}},\ \bibinfo {pages} {032110} (\bibinfo
  {year} {2018})}\BibitemShut {NoStop}%
\bibitem [{\citenamefont {{T. Oka and S. Kitamura}}(2019)}]{Floquet_review}%
  \BibitemOpen
  \bibfield  {author} {\bibinfo {author} {\bibnamefont {{T. Oka and S.
  Kitamura}}},\ }\href@noop {} {\bibfield  {journal} {\bibinfo  {journal} {Ann.
  Rev. Condens. Matter Phys.}\ ,\ \bibinfo {pages} {10:387}} (\bibinfo {year}
  {2019})}\BibitemShut {NoStop}%
\bibitem [{\citenamefont {{L. D'Alessio and A.
  Polkovnikov}}(2013)}]{D'Alessio}%
  \BibitemOpen
  \bibfield  {author} {\bibinfo {author} {\bibnamefont {{L. D'Alessio and A.
  Polkovnikov}}},\ }\href@noop {} {\bibfield  {journal} {\bibinfo  {journal}
  {Ann. Phys.}\ }\textbf {\bibinfo {volume} {333}},\ \bibinfo {pages} {19}
  (\bibinfo {year} {2013})}\BibitemShut {NoStop}%
\bibitem [{\citenamefont {{P. Ponte, A. Chandran, Z. Papic, and D. A.
  Abanin}}(2015)}]{Ponte}%
  \BibitemOpen
  \bibfield  {author} {\bibinfo {author} {\bibnamefont {{P. Ponte, A. Chandran,
  Z. Papic, and D. A. Abanin}}},\ }\href@noop {} {\bibfield  {journal}
  {\bibinfo  {journal} {Ann. Phys.}\ }\textbf {\bibinfo {volume} {353}},\
  \bibinfo {pages} {196} (\bibinfo {year} {2015})}\BibitemShut {NoStop}%
\bibitem [{\citenamefont {{L. Zhang, H. Kim, and D. A.
  Huse}}(2015)}]{Zhang_Floquet}%
  \BibitemOpen
  \bibfield  {author} {\bibinfo {author} {\bibnamefont {{L. Zhang, H. Kim, and
  D. A. Huse}}},\ }\href@noop {} {\bibfield  {journal} {\bibinfo  {journal}
  {Phys.\ Rev.\ E}\ }\textbf {\bibinfo {volume} {91}},\ \bibinfo {pages}
  {062128} (\bibinfo {year} {2015})}\BibitemShut {NoStop}%
\bibitem [{\citenamefont {{N. Regnault, and R. Nandkishore}}(2016)}]{Regnault}%
  \BibitemOpen
  \bibfield  {author} {\bibinfo {author} {\bibnamefont {{N. Regnault, and R.
  Nandkishore}}},\ }\href@noop {} {\bibfield  {journal} {\bibinfo  {journal}
  {Phys.\ Rev.\ B}\ }\textbf {\bibinfo {volume} {93}},\ \bibinfo {pages}
  {104203} (\bibinfo {year} {2016})}\BibitemShut {NoStop}%
\bibitem [{\citenamefont {{D. J. Luitz, R. Moessner, S. L. Sondhi, and V.
  Khemani}}(2020)}]{Luitz_Floquet}%
  \BibitemOpen
  \bibfield  {author} {\bibinfo {author} {\bibnamefont {{D. J. Luitz, R.
  Moessner, S. L. Sondhi, and V. Khemani}}},\ }\href@noop {} {\bibfield
  {journal} {\bibinfo  {journal} {Phys.\ Rev.\ X}\ }\textbf {\bibinfo {volume}
  {10}},\ \bibinfo {pages} {021046} (\bibinfo {year} {2020})}\BibitemShut
  {NoStop}%
\end{thebibliography}%

\end{document}